\documentclass{emulateapj}
\lefthead{Gaudi et al.} \righthead{NEARBY MICROLENSING EVENT}

\newcommand{\be}{\begin{equation}}
\newcommand{\ee}{\end{equation}}
\newcommand{\bea}{\begin{eqnarray}}
\newcommand{\eea}{\end{eqnarray}}

\begin{document}

\title{Discovery of a Very Bright, Nearby Gravitational Microlensing Event}
\author{ B. Scott Gaudi\altaffilmark{1}, Joseph Patterson\altaffilmark{2}, David S. Spiegel\altaffilmark{2}, 
Thomas Krajci\altaffilmark{3},
R. Koff\altaffilmark{4}, G. Pojma{\'n}ski\altaffilmark{5},  Subo Dong\altaffilmark{1}, 
Andrew Gould\altaffilmark{1}, Jose L. Prieto\altaffilmark{1}, Cullen H. Blake\altaffilmark{6}, Peter W. A. Roming\altaffilmark{7},
 David P. Bennett\altaffilmark{8},  Joshua S. Bloom\altaffilmark{9,10}, David Boyd\altaffilmark{11},  Michael E.\ Eyler\altaffilmark{12},
Pierre de Ponthi{\`e}re\altaffilmark{13},
N. Mirabal\altaffilmark{2}, Christopher W.\ Morgan\altaffilmark{1,12},  Ronald R. Remillard\altaffilmark{14}, T. Vanmunster\altaffilmark{15}, 
R.\ Mark Wagner\altaffilmark{16},
Linda C. Watson\altaffilmark{1}}

\altaffiltext{1}{Department of Astronomy, The Ohio State University, 140 W.\ 18th Ave., Columbus, OH 43210, USA}
\altaffiltext{2}{Department of Astronomy, Columbia University, 550 West 120th Street, New York, NY 10027, USA}
\altaffiltext{3}{Center for Backyard Astrophysics (New Mexico), 9605 Goldenrod Circle, Albuquerque, NM 87116, USA}
\altaffiltext{4}{Center for Backyard Astrophysics (Colorado), Antelope Hills Observatory, 980 Anteleope Drive West, Bennett, CO 80102, USA}
\altaffiltext{5}{Warsaw University Astronomical Observatory, Al. Ujazdowskie 4, 00-478 Warszawa, Poland}
\altaffiltext{6}{Harvard-Smithsonian Center for Astrophysics, 60 Garden St., Cambridge, MA 02138, USA}
\altaffiltext{7}{Department of Astronomy and Astrophysics, Pennsylvania State University, 525 Davey Lab, University Park, PA 16802, USA}
\altaffiltext{8}{Department of Physics, University of Notre Dame, IN 46556, USA}
\altaffiltext{9}{Department of Astronomy, 601 Campbell Hall, University of California, Berkeley, CA 94720-3411, USA}
\altaffiltext{10}{Sloan Research Fellow}
\altaffiltext{11}{Center for Backyard Astrophysics (England), 5 Silver Lane, West Challow, Wantage, OX12 9TX, UK}
\altaffiltext{12}{Department of Physics, United States Naval Academy, 572C Holloway Road, Annapolis, MD 21402, USA}
\altaffiltext{13}{Center for Backyard Astrophysics (Lesve), 15 rue Pr{\'e} Mathy, 5170 Lesve (Profondeville), Belgium}
\altaffiltext{14}{MIT Kavli Institute for Astrophysics and Space Research, Massachusetts Institute of Technology, 77 Massachusetts Avenue, Cambridge, MA 02139-4307, USA}
\altaffiltext{15}{Center for Backyard Astrophysics (Belgium), Belgium Observatory, Walhostraat 1A, B-3401 Landen, Belgium}
\altaffiltext{16}{Large Binocular Telescope Observatory, University of Arizona, 933 North Cherry Ave., Tucson, AZ 85721, USA}

\begin{abstract}
We report the
serendipitous detection of a very bright, very nearby microlensing event. In late
October 2006, an otherwise unremarkable A0 star at a distance
$\sim 1~{\rm kpc}$ (GSC 3656-1328) brightened achromatically by a factor
of nearly 40 over the span of several days and then decayed in an
apparently symmetrical way.  We present a
light curve of the event based on optical photometry from the Center for Backyard Astrophysics
and the All Sky Automated Survey, as well as near-infrared photometry
from the Peters Automated Infrared Imaging Telescope.  This light curve is well-fit by a generic microlensing
model. We also report optical spectra and {\it Swift} X-ray and UV observations
that are consistent with the microlensing interpretation.  We discuss
and reject alternative explanations for this variability.  The lens
star is probably a low-mass star or brown dwarf,
with a relatively high proper motion of $\ga 20~{\rm mas~yr^{-1}}$, and may
be visible using precise optical/infrared imaging taken several years
from now.  A modest, all-sky survey telescope
could detect $\sim 10$ such events per year, which would enable 
searches for very low-mass planetary companions to relatively nearby stars.
\end{abstract}

\section{Introduction}

\bigskip

At the very moment in 1936 that he introduced\footnote{There is some
confusion as to who first worked out the basic concepts of
gravitational microlensing.  Indeed, \citet{eddington20} and
\citet{chwolson24} both discussed the possibility in the 1920s.
However further research has shown that Einstein had already worked
out the basic formalism of microlensing in 1912 \citep{renn97}, modulo
the famous ``factor of 2'' increase in the deflection of light that he
only discovered when he introduced general relativity 4 years later.
In fact, \citet{soldner04} derived the classical value for the deflection
of light by a massive body over 100 years before Einstein, although he did
not consider the associated magnification of the source by the lens. 
See \citet{lenses} for a more thorough discussion of the history
of gravitational lensing.}
the concept of gravitational microlensing of one star by another
closely aligned along its line of sight, Einstein famously dismissed
its practical significance.  Noting that the characteristic scale
(what we now call the ``Einstein radius'') was extremely small, he
concluded that ``there is no great chance of observing this
phenomenon, even if dazzling by the light of the much nearer star is
disregarded.'' \citep{einstein36}.  
It is easily shown that the
optical depth to microlensing (the probability that any given star
lies projected within the Einstein radius of another) is only about
$\tau\sim 10^{-8}$ among the $V\leq 12$ stars that were typically
cataloged in Einstein's day.  As there are only a few million such
stars altogether, the probability that any of these are microlensed at
any given time is much less than one \citep{nemiroff98}.  Such a
calculation may have influenced Einstein to resist the determined
efforts by Hungarian engineer R.\ W.\ Mandl to get Einstein to publish
his microlensing formulae and perhaps also to compose a letter to the
editor of {\it Science} to ``thank you for your cooperation with the
little publication, which Mister Mandl squeezed out of me.  It is of
little value, but it makes the poor guy happy'' \citep{renn97}.

Since Einstein's 1936 article, several authors have attempted to resurrect
the idea of microlensing (e.g., \citealt{liebes64,refsdal64}). However,
microlensing experiments did not actually get underway until the early
1990s \citep{alcock93,aubourg93,udalski93}.  These experiments were
motivated both by the suggestion of \citet{pac86} to simultaneously
monitor millions of stars in the dense star fields of nearby galaxies
and by contemporaneous advances in technology that made such
experiments feasible.

To date, several thousand microlensing events have been discovered
toward several lines of sight, including the Large and Small
Magellenic Clouds \citep{alcock97,palanque98,alcock00}, as well as M31
\citep{ph02,dejong04,uglesich04,calchi05}.  However, the vast majority of
events have been detected toward the Galactic bulge
\citep{udalski00,thomas05,hamadache06} or fields in the Galactic
plane relatively close to the bulge \citep{derue01}.  The source stars of these
events have generally been relatively faint, $V \ga 15$. In these cases,
the optical depth is of order $10^{-6}$, i.e.\ $100$ times higher than
for the local stars that Einstein would have considered because the
targets are about 10 times farther away. The larger distance makes the
area of the Einstein ring about 10 times bigger and increases the
column density of potential lenses by another factor of 10.  

In the intervening years since the first microlensing events were
discovered, a few authors have revisited the idea of detecting bright
and/or nearby 
microlensing events.  \citet{colley95} argued that microlensing events
visible to the naked eye are exceedingly rare, occurring at a rate of
one per 40,000 yr for lensing by known stars. Microlensing of
fainter stars is obviously more common; the event rate for stars with
$V\la 15$ is roughly one per year \citep{nemiroff98,han07}.  This
calculation led \citet{nemiroff98} and \citet{nemiroff99} to point out
that only small-aperture instruments are required to discover these
brightest microlensing events, and that monitoring the entire sky down
to $V \sim 15$ on relatively short timescales would soon be feasible.
\citet{distefano05} considered in detail a related idea of
``mesolensing'': microlensing of background stars by nearly lenses
with large angular Einstein rings and large proper motions.
Mesolensing can be used to study the properties of nearby stars, and in particular 
their planetary companions \citep{distefano07,distefano08,dinight08}.  See 
\S\ref{sec:disc} for further discussion of the potential of
microlensing to discover and study nearby planetary systems.

Although, given the typical source distances of $\sim 8-50~\rm kpc$ for
microlensing searches toward the Galactic bulge and the Magellanic
Clouds, the majority of the microlenses giving rise to observed events
have been located at distances of a kiloparsec or more, there is
nevertheless a low-probability tail of more nearby microlenses
\citep{gould94,distefano05}.  Indeed, there is one event for which the
lens has been robustly located to within a kiloparsec of the Sun,
MACHO-LMC-5 \citep{alcock01,gould04}.  Another event, EROS2-LMC-8, is also
likely due to a nearby microlens, although this has yet to be
confirmed with follow-up observations (J.P. Beaulieu 2008, private
communication; see also \citealt{tisserand04} and
\citealt{tisserand07}).  Thus, nearby lenses can be uncovered and
studied in ongoing surveys.  However, these surveys are not very
efficient at discovering nearby microlenses; furthermore, the
source stars of these events are necessarily faint, making detailed monitoring
and follow-up observations  difficult.  

Here we report on a microlensing event of the type Einstein
believed would never be observed: a magnification $A=40$ event of the 
$V\sim 11.4$ A-type star GSC 3656-1328, located about 1 kpc from the Sun in the disk of the
Milky Way. Although such  events are indeed quite rare, microlensing
of somewhat fainter nearby stars occurs with reasonable frequency.
While this event was found serendipitously, we argue that with
recent technological advances it is now feasible to monitor the sky to
deliberately and routinely detect these ``domestic'' microlensing events.
We propose a telescope design to accomplish this, and we argue that
it is possible to build two copies of such a telescope that could
monitor the majority of the sky down to $V\sim 16$ at modest cost.
Monitoring of the discovered microlensing events would enable a novel
method to detect nearby planets, allow a search for dark objects in
the Milky Way disk, and permit several days advance warning for potentially
hazardous asteroid impacts.

\citet{fukui07} also report on observations of the transient event in
GSC 3656-1328.  A subset of the data presented here is in common
with theirs; however, the analyses were done completely independently.
They also conclude that the brightening seen in GSC 3656-1328 is
likely due to microlensing.

\section{Observations}

On 2006 October 31, A.\ Tago announced
a sudden brightening in the close vicinity of GSC 3656-1328 \citep{nakano06,nakano06b}, a $V=11.4$
A0V-A1V star\footnote{ We summarize the
properties of GSC 3656-1328 in Appendix \ref{app:source}.} $\sim 8^\circ$ from the Galactic plane
in Cassiopeia with a distance of $\sim 1~{\rm kpc}$. 
Confirmation followed a few hours later in the variable-star
newsgroups; this established that the star was in fact the source of
the brightening, and roughly measured the amount and timescale of the
brightening (to V=7.5, rising in about a week).

We began our photometry with the small telescopes of the Center for
Backyard Astrophysics (CBA: \citealt{skillman93}) on
November 1 (=JD$'$=JD-2450000=4040.), and found the star at $V=8.9$, 
falling rapidly and smoothly, with no additional variability at the
few percent level.  In the next few days, we obtained time-series and
multicolor photometry with CBA telescopes, spectroscopy with the MDM
2.4 m telescope, and a pointed 5000 s X-ray observation with
{\it Swift} \citep{gehrels04} using the XRT \citep{burrows05}; we also
searched for X-ray outbursts over the 10-yr lifetime of the All-Sky
Monitor aboard {\it RXTE}.  The basic conclusion from all of these
observations was simple: the star had brightened by 4 mag, but
with no change in spectrum or color ($B-V=0.2$), no flickering, and no
discernible X-ray signature (with a {\it Swift} upper limit in the 0.5-10
keV band of $<4\times 10^{-13}~{\rm ergs~cm^{-2}~s^{-1}}$).  A search
of 400 photographic plates during 1964-1994 showed no variability in the
star \citep{samus06}, and we found no X-ray outbursts from this
position.
                           
This seemed unlike any known type of intrinsic variable star, and
several reports suggested instead that GSC 3656-1328 may have
microlensed by an intervening passing star
\citep{mikolajewski06a,mikolajewski06b,spiegel06}.  After 2 weeks of
data, our preliminary fit of the light curve to a microlens model (see
\S\ref{sec:micro}) seemed promising.  However, there were only a few
very uncertain measurements (with typical errors of $\pm 0.4$ mag) before the peak,
and identifying microlensing events based on falling-only light curves
can be problematic \citep{smith03,afonso06}.  Given the low {\it a
priori} likelihood for such a nearby star to be microlensed, we eagerly
sought additional data to confirm (or refute) the microlensing
interpretation.  In particular, we sought data before the apparent
peak on October 31, as well as data covering a wide baseline in
wavelength, in order to ascertain whether the variation was both symmetric
about the peak and achromatic, as would generally be expected
for a short microlensing event of an isolated star.
 
We were fortunate to find 15 images of the field in the $V$ and $I$
bands in the test runs of the northern (Hawaii) station of the All-Sky
Automated Survey (ASAS; the southern station is described by
\citealt{pojmanski04}).  These images have errors of $\pm 0.04$ mag,
cover (athough sparsely) the rise of the event, and fortuitously
include one $I$-band measurement obtained almost exactly at the peak,
judged by a microlensing fit to the remainder of the data set, as
described below.  Since ASAS-North had just begun operation, this was
mighty lucky.

As for wavelength coverage, we supplemented the CBA monitoring with
four epochs of UV observations (on JD$'=4042$, 4055, 4057, and 4067)
using the UVOT camera on {\it Swift} \citep{roming05}, and a long program of
infrared monitoring from the Peters Automated Infrared Imaging
Telescope (PAIRITEL, \citealt{bloom06}) on Mount Hopkins.  PAIRITEL is a 1.3m telescope
equipped with the former 2MASS \citep{skrutskie06} camera that
simultaneously images the near infrared $JHK_{s}$ bands.  We obtained
$\sim 1500$ 7.8 s images in each band nearly every night from program
start (JD$'=4044$) until the event returned to baseline.  Since the
variations were smooth, we binned the images into 51 separate epochs,
and the magnitude of the source at each epoch was determined using
differential aperture photometry against a set of reference stars from
the 2MASS catalog.  The {\it Swift} UV data from this bright source were
strongly affected by coincidence losses, and to minimize these we used
only the M2 band (centered on 2400\AA) and estimated the flux of the
source by summing in an annulus centered on the source, with an inner
radius of $12''$ and an outer radius of $20''$.

\begin{figure}[ht*]
\epsscale{1.0}
\plotone{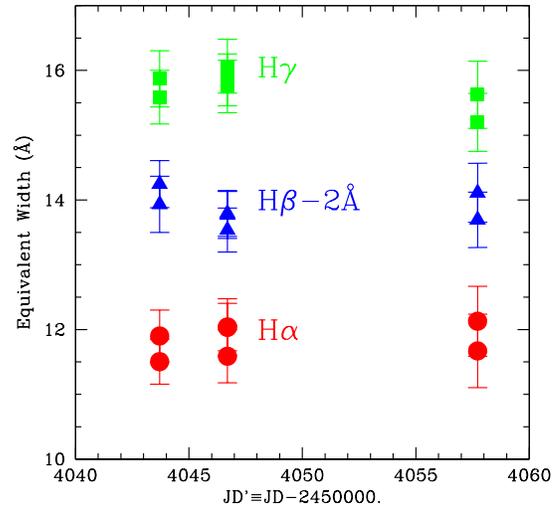}
\caption{
EW of the H$\alpha$ (red circles), H$\beta$ (blue
triangles), and H$\gamma$ (green circles) lines of GSC 3656-1328 as a
function of JD-2450000.  The measurements for H$\beta$ have been
offset by 2\AA for clarity, as indicated.  The broadband optical
photometry in Fig.\ \ref{fig:lcurve} demonstrates that during the time
span between the first and last spectroscopic measurements, the
continuum flux of GSC 3656-1328 decreased by a factor of $\sim 2$,
whereas the EWs of H$\alpha$, H$\beta$, and H$\gamma$ remained
constant to within the uncertainties.
\label{fig:ew}
}
\end{figure}

We also obtained seven optical spectra using the CCDS instrument on the
MDM 2.4m telescope on Kitt Peak, two near JD$'\sim 4043.7$, three
near $\sim 4046.7$, and two near $\sim 4057.7$. The spectra 
cover $4000-6800$\AA, with a
resolution of $\sim 15$\AA\ FWHM, as measured from arc-lamp
lines. They reveal an unremarkable early A star, showing prominent
resolved (FWHM$\sim 25$\AA) Balmer lines.  We measured the equivalent
widths (EWs) of H$\alpha$, H$\beta$, and H$\gamma$ in each spectrum by
integrating over a window of $\pm 60$\AA, fitting a third-order
polynomial to estimate the local continuum.  We estimated the
uncertainties in the EW measurements by generating a series of mock
spectra with identical noise properties, and measuring the EWs in
these spectra in the same manner as the actual data.  See Figure
\ref{fig:ew}.

\begin{figure}[ht*]
\epsscale{1.2}
\plotone{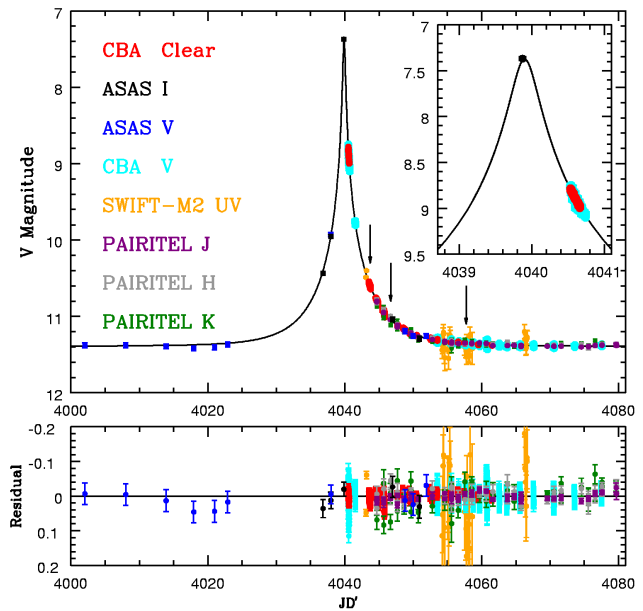}
\caption{Data and residuals from a microlensing model fit to the UV,
optical, and IR photometry of GSC 3656-1328.  (Top panel) The points
with error bars show photometry from CBA clear (red) and $V$ (cyan),
ASAS $I$ (black) and $V$(blue), {\it Swift} M2 (orange), and PAIRITEL $J$
(purple), $H$ (grey), and $K_s$ (green), as a function of JD$'=$JD-2450000.  All data except for the ASAS $I$ and $V$ bands have been
aligned to the CBA $V$-band by subtracting a constant magnitude
offset.  For the ASAS data, the additional contribution from
unresolved neighboring stars has been removed using the microlensing
model fit. See text for details.  The solid line shows the best-fit
microlensing model.  The arrows show the epochs of the spectroscopic
measurements.  The inset shows a detail near the peak of the event. (Bottom panel) The residuals from the best-fit
microlensing model.  The achromatic and symmetric variability
of GSC 3656-1328 is well-fit by a microlensing model.
\label{fig:lcurve}
}
\end{figure}

\section{Microlensing Interpretation}\label{sec:micro}

Figure \ref{fig:lcurve} shows the collected UV, optical, and near-IR
photometry of GSC 3656-1328 during the transient variability.  For all of
these data except those from the CBA, the photometric uncertainties are
due to the photon noise from the source and
comparison stars.  In the case of the CBA data sets, we estimated the uncertainties
to be equal to the rms variability about a linear fit to the time-series data on the first night (JD$'=4040$).  
We align all of the bands to the CBA $V$-band by subtracting a
constant magnitude offset, except for the ASAS data, which due to the
larger point-spread function (PSF) and photometric aperture include an additional
contribution from unresolved neighboring stars.  In this case, we
remove this unresolved (``blended'') light using the microlensing
model fit, as described below.  
The baseline magnitude of the microlensing source determined in this way is
consistent with that measured directly from higher resolution images.  
As is evident from Figure \ref{fig:lcurve}, the variability is
essentially achromatic (i.e.\ the color of the event is constant) over
nearly a decade in wavelength.  Furthermore, although the sampling on
the rising side of the event is sparse, the variability is apparently
symmetric about the time of peak brightness near JD$'\simeq 4040$.

Our spectra show no evidence for any emission features, and are
completely consistent with that of a normal A0V star.  Furthermore,
there is no evidence for any change in the spectrum of the star
between JD$'\simeq 4043.7$ and $4057.7$, during which time the
continuum flux decreased by a factor of $\sim 2$.  This can be
quantified by the EW of the Balmer lines, which were
constant to within the measurement uncertainties (typically $2\%-4\%$)
during this period (see Fig.\ \ref{fig:ew}).  Reports 
of optical spectra taken by other groups, which span a broader range of epochs, 
apparently confirm this lack of spectral evolution \citep{munari06,mikolajewski06a,mikolajewski06b}. 

The target-of-opportunity observation by the {\it Swift} satellite showed no
detectable X-ray emission, as would be expected from outbursting
variables.

These characteristics of the variability in GSC 3656-1328, namely the
precise light curve shape including the apparent symmetry about the peak,
achromaticity, and lack of emission features or changes in the
absorption features, are expected for microlensing but would be very
unusual for outbursting variable stars, which generally change
temperature (and therefore color) during explosive outbursts.
Essentially, what is required to reproduce the 
type of variability seen in GSC 3656-1328 is for the angular size of the star to
change by a factor of $\sim \sqrt{40}\sim 6$ while the temperature remains constant
to $\la 5\%$.  The only way for this to occur is for the apparent area
of the star to change, as occurs in a microlensing event.  Occultation
of the star due to, e.g., an eclipsing binary companion can also make such a
change in the apparent area, but obviously in this case one would
expect a dimming, not the brightening that is observed here.  

When microlensing surveys were first being planned, a major concern
was the potential contamination from intrinsic stellar variability.
As a result, a number of studies addressed the question of whether
there exists a class of variable stars whose variability might be
misinterpreted as microlensing.  At least two types of potential
contaminants were identified.  

``Blue bumpers'' are blue main-sequence stars whose fluxes remain
constant for long periods of time, but occasionally undergo
brightenings (``bumps'') that are approximately symmetric about the
peak, and have durations that are consistent with the expectations for
microlensing events \citep{cook95,alcock96,alcock00}.  However, closer
inspection reveals that these variables have properties that can be
used to exclude them as the explanation for the variability seen in GSC
3656-1328.  First, many or perhaps all of these variables are Be-type stars,
exhibiting Balmer line emission \citep{cook95}.  Second, the
brightenings are exclusively low amplitude, with peak brightenings of
less than 1.5 times the baseline flux \citep{alcock00}.  Finally,
detailed light curves typically show slight asymmetries
\citep{cook95,alcock00}.

At least some dwarf nova (DN) outbursts are characterized by approximately
symmetric brightenings with amplitudes and durations that are
consistent with that seen in GSC 3656-132
\citep{dellavale94,beaulieu95,dellavale96}. However,
it is unlikely that the GSC 3656-132
variability is due to a DN outburst.  First, spectra of DNs typically exhibit
H or He emission \citep{dellavale94,beaulieu95}.  Second, while the light
curves of novae have properties that are grossly similar to
microlensing events, they do not follow the standard \citet{pac86}
form at the $\sim 2\%$ level with which the GSC
3656-1328 variability has been measured (e.g.,
\citealt{smith03,afonso06}).  Finally, the interval between outbursts of DNs is
known to be directly related to the strength of the outburst, such
that larger outbursts generally have longer intervals between
outbursts \citep{smak84,vp85}. DN outbursts with amplitudes similar to the
brightening seen in GSC 3656-132 would be expected to have average
recurrence times of tens to hundreds of days \citep{vp85}.  It 
seems 
unlikely that the kinds of observations that originally discovered the
variability discussed here (e.g., \citealt{nakano06,nakano06b}) 
would have missed previous outbursts; furthermore, there is no
evidence for such outbursts in the photographic plate observations
taken during 1964-1994 \citep{samus06}.
Regardless, continued monitoring of GSC 3656-1328 would certainly
allow one to rule out the DN hypothesis definitively.

It is interesting to note that the source stars of the first two EROS
microlensing event candidates \citep{aubourg93} are both early-type
main-sequence stars. Spectroscopic observations showed that the source
of EROS-LMC-1 is a Be star with Balmer emission lines, whereas the
source of EROS-LMC-2 is a seemingly normal A0 main-sequence star
with no apparent emission lines \citep{beaulieu95}.  However,
EROS-LMC-2 also exhibits periodic variability with an amplitude of
$\sim 0.5$ mag and a period of $\sim 2.8$ days, indicative of an
eclipsing binary \citep{ansari95}.  In both cases, continued
photometric monitoring of the source stars revealed additional
brightenings many years later, with amplitudes and timescales 
similar to those of the original events \citep{lasserre00,tisserand07}, thus
excluding the microlensing interpretation of the variability.

Given the evidence in favor of microlensing, we first test whether the UV, optical,
and IR data can be acceptably fit to a simple microlensing model.  This
model has as parameters  the Einstein timescale $t_{\rm E}$, 
the impact parameter (closest source-lens approach in units of the
angular Einstein radius $\theta_{\rm E}$) $u_0$, and the time-of-maximum 
magnification $t_0$.  In addition, we fit for the baseline
flux of the lensed source 
for each of the eight separate observatory/filter combinations.
Since we expect any light in the PSF to
be completely dominated by the bright source, we do not allow for any flux that is blended
with the source but is not microlensed, with the exception of
the ASAS data, which are known to contain light from unresolved stars in the
photometric aperture.   
Thus, this model has $3+7+2=12$ parameters.  This best-fit model is shown in Figure \ref{fig:lcurve}. 
We find that the microlensing model provides a reasonable
fit to the data: the $\chi^2/{\rm dof}$ for the individual data sets ranges
from $\sim 0.8$ for 2802 data points for the CBA $V$ data set to $\sim
2.6$ for 34 data points in the worst case of the {\it Swift} UVM2 dataset.
In the latter case, the data may be somewhat compromised by the fact that the
peak of the source PSF was affected by coincidence losses in the first two exposures, although we
attempted to circumvent this difficulty by using annular apertures. The
remaining statistically significant deviations from $\chi^2/{\rm dof}$
of unity can be traced to large systematic outliers and  
well-known correlated systematic errors in
the photometry.  In addition, as we discuss in detail below, the PAIRITEL data show 
some evidence for a small but non-zero component of blended light which may be due to
the lens itself.  We therefore conclude that the microlensing model provides
a good representation of the data.

Accepting the microlensing interpretation, we then refit the
microlensing model.  We use an iterative procedure to
remove large outliers from the CBA data ($\ga 3\sigma$ for CBA $V$ and
$\ga 3.5\sigma$ for CBA clear) and re-normalize the uncertainties in
each data set by a constant factor to force $\chi^2/{\rm dof}=1$,
except for the UV data, for which we instead add a constant 0.03
mag uncertainty in quadrature.  To provide limits on the
magnitude of any contribution of light from the lens itself, we now
allow for two parameters for each of the eight separate
observatory/filter combinations: the flux of the lensed source and a
free term for any flux that is blended with the source but is not
microlensed.  This model has $3+2\times 8=19$ parameters.  We find
$t_{\rm E}=7.19\pm 0.03$ days, $u_0=0.0237\pm 0.0007$, and
$t_0=4039.899 \pm 0.003$.    This
timescale is unexpectedly short; for a typical lens
velocity  relative to the observer-source line of sight of $v_\perp \sim 70~{\rm km~s^{-1}}$ , a lens mass of $M \sim
0.3~M_\odot$, and a lens distance of $D_L\sim 500~{\rm pc}$, one would
expect $t_{\rm E}\sim 20~{\rm days}$.   This
implies that the lens is moving fast, is of low mass, 
is very close to the source or observer, or some combination
of these. 

In order to provide more quantitative constraints on the nature of the
lens, we perform a Bayesian analysis.  We adopt priors for the
parameters of microlensing events expected toward the GSC 3656-1328
line of sight generated from a Galactic model which includes double
exponential thin and thick disks, an exponential distribution of dust
in the vertical direction, and a mass function of lenses including
stars, brown dwarfs, and remnants.  We include constraints derived from the 
measured proper
motion of the source, as well as constraints from the analysis of the
light curve. These latter constraints include the measurement of the
timescale of the event, as well as limits on the angular size of the
source in units of the Einstein ring radius, and the flux of the lens.
Details of the Bayesian analysis, including information about the
model assumptions and parameters, as well as the observational
constraints, are provided in Appendix \ref{app:lens}.

Figure \ref{fig:probs} shows the Bayesian probability densities for the
properties of the lens star giving rise to the microlensing event seen
in GSC 3656-1328.  Shown are the results for the lens mass, distance,
relative lens-source proper motion $\mu$, Einstein ring radius $R_{\rm
E}$, and the $V,J,H,K_{s}$ magnitudes of the lens assuming it
is a main-sequence star.  The median and 68.3\% confidence intervals
are $\log(M/M_{\odot})=-1.06_{-0.35}^{+0.39}$,
$D_L=420_{-250}^{+320}~{\rm pc}$, $\mu=43_{-21}^{+63}~{\rm mas~yr^{-1}}$
and $R_{\rm E}=0.35_{-0.10}^{+0.17}~{\rm AU}$.  There is an $\sim 49\%$
probability that the lens is a main-sequence star, an $\sim 46\%$
probability that the lens is a brown dwarf, and an $\sim 5\%$
probability that the lens is a white dwarf.  The probability that the
lens is a neutron star or black hole is negligible.  The velocity of the lens relative to the observer-source line of sight is $v_\perp = 84_{-25}^{+41}~{\rm km~s^{-1}}$, indicating
that it is probably a member of the thick disk.  Thus the most likely
scenario for the lens is that it is a low-mass star or brown dwarf in the thick disk,
with a mass near the hydrogen burning limit.  The proper motion is
likely to be quite high, with $\mu \ga 16~{\rm mas~yr^{-1}}$ at the 95\%
confidence level.  The apparent magnitudes of the lens, assuming it is
a main sequence star, are $V=23.6_{-3.8}^{+3.9}$,
$J=17.3_{-2.5}^{+1.9}$, $H=16.6_{-2.4}^{+1.8}$, and
$K_s=16.2_{-2.4}^{+1.7}$.  Therefore, light from the lens may be
directly detectable in a few years, when it has separated from the
source, using high-resolution, near-infrared imaging.

\begin{figure}[ht*]
\epsscale{1.6}
\plotone{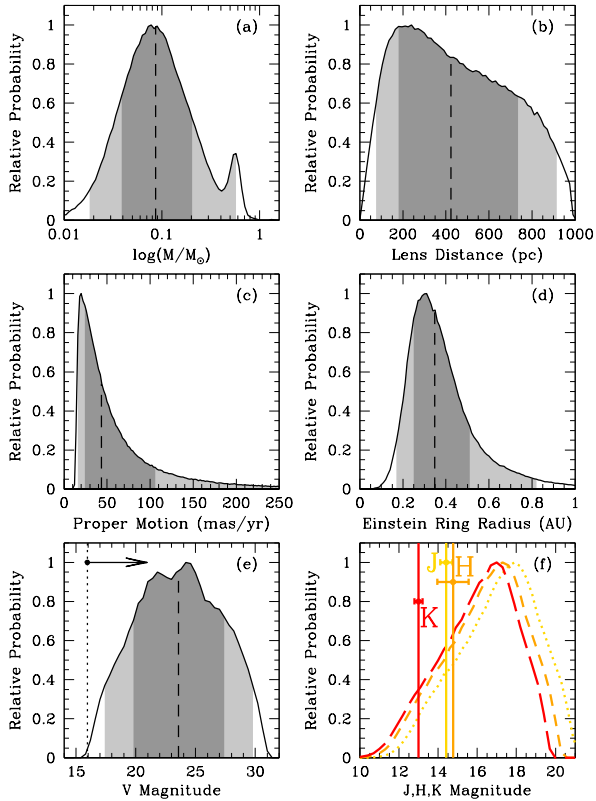}
\caption{
Bayesian probability densities for the properties of the lens
star giving rise to the microlensing event seen in GSC 3656-1328.
These distributions are derived assuming priors derived from 
models of the mass, velocity, and density distributions of stars in the thin and thick
disks, and include constraints on the proper motion of the source and
timescale of the microlensing event, as well as limits on the proper motion
and $V$-band magnitude of the lens as derived from the light curve.  
See text.  The panels show the probabilities for (a) the lens mass
(b) the lens distance, (c) the lens proper motion, (d)
the Einstein ring radius of the lens, (e) the $V$ magnitude
of the lens, and (f) the $J$ (dotted), $H$ (short dashed), and $K_s$ (long dashed) 
magnitude of the lens.  In panels (a)-(e), the dashed line shows the medians of the distributions,
and the dark and light shaded regions encompass the 68.3\% and 95.4\%
confidence intervals, respectively.  In panel (e) the dotted
line shows the 95\% c.l.\ upper limit on the $V$-band flux
from the lens.  In panel (f), 
the solid lines show the measured blend fluxes from the $JHK_s$ light curves.
The points show the same along with associated uncertainties; the abscissa values
are arbitrary. 
\label{fig:probs}
}
\end{figure}

\begin{figure}[ht*]
\epsscale{1.6}
\plotone{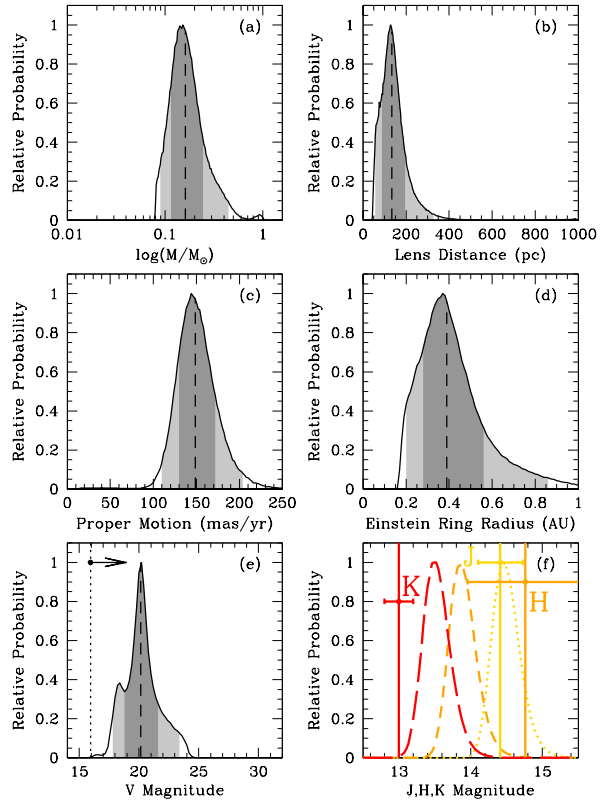}
\caption{
Same as Figure \ref{fig:probs}, but assuming that the blend
fluxes measured in the $JHK_s$ light curves are real
and correspond to flux from the lens.  This
constrains the mass and distance to the lens, and
implies that the lens is relatively nearby with a large
proper motion of $\mu \ga 100~{\rm mas~yr^{-1}}$.
Note the greatly compressed scale in panel (f) as compared to Figure \ref{fig:probs}.
\label{fig:probs_ir}
}
\end{figure}

Given that the source is quite luminous, and the lens likely to be of
relatively low mass, one would generally expect any blended light due
to the microlens to be very small by comparison. Indeed the measured
blend flux values for the CBA $I$, CBA $V$, and {\it Swift} M2 data sets are
all consistent with zero and less than 2\% of the source flux.  For
the ASAS $I$ and $V$ datasets, the blending is significant, but as
discussed previously, the ASAS photometric aperture is known to
contain light from nearby stars that are unresolved by ASAS but
resolved by the CBA data.  Surprisingly, we find evidence for
significant blended light in the $J$ and $K_s$ bands, and marginal
evidence in the $H$ band.  Specifically, we find blend magnitudes of
$(J)_B=14.41\pm0.31$, $(H)_B=14.76\pm 0.81$, and $(K_s)_B=13.00\pm 0.20$.
The color and magnitude of this blended light are roughly consistent
with that of a mid-to-late $M$ dwarf located $\sim 100~{\rm pc}$ away.
Although it is {\it a priori} unlikely that the lens would be
sufficiently close that its light could be detectable against such a
bright source, the short timescale of the event already argues for a
somewhat more nearby lens, making the detection of blended light more
plausible.  Indeed, as can be seen from Figure \ref{fig:probs}, even
ignoring any potential constraints from the $JHK_s$ flux of the lens,
there is a small but non-negligible probability for the flux from the
lens to be as large as the measured blended light.  

If we assume that the blended light is indeed due to the lens,
we can include this information in the Bayesian analysis
to provide much stronger constraints on the properties of the lens \citep{bennett06}.
Figure \ref{fig:probs_ir} show the resulting probability densities.
We find that the mass, distance, and proper motion are more
tightly constrained, $\log(M/M_{\odot})=-0.79_{-0.15}^{+0.19}$,
$D_L=130_{-48}^{+62}~{\rm pc}$, and $\mu=150_{-20}^{+24}~{\rm mas~yr^{-1}}$,
whereas the constraints on the Einstein ring radius are quite similar 
($R_{\rm E}=0.39_{-0.12}^{+0.17}~{\rm AU}$).  
The lens apparent magnitudes are $V=20.2_{-1.4}^{+1.5}$,
$J=14.5_{-0.18}^{+0.21}$, $H=13.9_{-0.17}^{+0.20}$, and
$K=13.5_{-0.18}^{+0.21}$.  We note that the median
posterior value of the $K_s$ apparent magnitude
differs substantially from the input constraint due to the strong
prior that the lens be more distant and less massive, and hence
fainter.   

If the blended light is real and indeed due to the lens,  
the lens must be fairly nearby
and have a high proper motion ($\mu \ga 110~{\rm mas~yr^{-1}}$ at
95\% confidence), and so should be resolved from the source 
in a few years.  However, there are caveats. The IR data 
in this event (as well as other data taken by PAIRITEL) show evidence
for low-level systematic errors at the few percent level that are correlated 
on the timescale of several days.  Furthermore, the early IR data were taken when
the source was sufficiently bright that non-linearity may be important. Either of
these systematic effects could easily give rise
to a spurious blending signal.  Regardless, the hypothesis that 
the lens is nearby should be testable in the near future with
high-resolution IR imaging.  A measurement 
of the lens proper motion and relative lens-source parallax, when combined with
the timescale of the microlensing event, would allow for a measurement
of the lens mass \citep{refsdal64,pac95}.

The rate $\Gamma$ of microlensing events of a single source lying at a 
distance $D$ by a uniform population of perturbers of number density $n$ is \citep{pac86}
\begin{equation}
\Gamma ={\pi\over 2}G^{1/2}n\langle m^{1/2}\rangle 
{\langle v_\perp\rangle \over c} D^{3/2},
\label{eqn:gamma}
\end{equation}
where $G$ is the gravitational constant, 
$\langle m^{1/2}\rangle$ is the mean square-root mass of the perturbers
and $\langle v_\perp\rangle$ is their mean velocity relative to the
observer-source line of sight.  Adopting $\langle
m^{1/2}\rangle=(0.5\,M_\odot)^{1/2}$, $\langle v_\perp\rangle =
55\,{\rm km\,s^{-1}}$, $n=0.1\,\rm pc^{-3}$, and $D=1\,\rm kpc$, this
yields $\Gamma = 0.043\,\rm Myr^{-1}$.  Since there are of order 2
million stars in the Tycho-2 catalog and noting that the source (GSC
3656-1328) is near the brightness limit of this catalog, one would
expect one microlensing event of a Tycho-2 star every $\sim 12$ yr.
This crude estimate is in rough agreement with the more detailed
calculations of \citet{han07}, who finds a 
rate of one event every $\sim 6$ yr for stars brighter than
$V=12$.  Thus, at first sight, the detection of a microlensing
event with a source star with $V\sim 12$ seems very plausible, given that amateur
and professional observers have been combing the skies for comets for
almost 40 years.  However, it is unlikely that this microlensing event
would have stimulated enough interest to generate the high quality
follow-up observations that made a convincing case for microlensing if
it had not been magnified by at least $A>10$.  Such events are a
factor $A^{-1} = 1/10$ less likely to occur.  Furthermore, the
fraction of the event duration when the source is magnified by $>A$ is also $\sim
A^{-1}$.  Taking these factors into account, we were probably lucky to observe
such a microlensing event, but perhaps not unreasonably so.

\section{Discussion}\label{sec:disc}

The above calculation and the more detailed study by \citet{han07} 
show that the event rate for sources 
$\sim 1\,$kpc would be quite small, even with a more thorough and
aggressive search procedure that detected essentially all lenses that
come within 1 Einstein radius ($A\sim 1.34$) of the source.  However,
if the search could be extended from 1 to 4 kpc, then the event rate
would be increased by $4^{7/2}=128$ times, to roughly 8 per year.
That is, we obtain $D^{3/2}$ from equation \ref{eqn:gamma} and $D^2$
from the larger volume probed (since viable targets are effectively
confined to the two-dimensionsal structure of the Galactic plane).  Such a survey
would enable a probe of the matter distribution near the Sun that was
equally sensitive to dark and luminous objects (see also \citealt{distefano05}).  It would also bring
microlensing planet searches, which have been successful at detecting
some novel planets against distant sources in the Galactic bulge
\citep{bond04,udalski05,beaulieu06,gould06}, nearer to home \citep{distefano07,distefano08,dinight08}.  

In order to provide an illustrative example of the planet discovery
potential of such nearby microlensing events, we determine what the
sensitivity of this microlensing event to planetary companions would
have been had it been discovered well before peak.  Specifically, we
create a fake data set and then  determine the planet detection
sensitivity of this data set using the methods outlined
in \citet{dong06}.  We assume that before JD$'=4034.0$, the event was sampled
at a rate of one
point per day.  For the time intervals of $4034.0 \le$JD$'\le 4037.0$
and $4046.0 \le$JD$'\le 4080.0$, we assume one point every 30
minutes, and over the peak, $4037.0 \le$HJD$'\le 4046.0$, we assume
one point per minute.  We assume a photometric uncertainty of 0.4\% for
each data point.  These assumptions are reasonable given the large
number of amateur and professional northern telescopes that are
available to follow these events, as well as the bright apparent
magnitude of the event.  We then determine the planet detection
sensitivity of this simulated data for planet/star mass ratios
of $q=10^{-3}$, $10^{-3.5}$, $10^{-4}$, $10^{-4.5}$, and $10^{-5}$.

The finite size of the source will begin to suppress perturbations for
mass ratios such that $\sqrt{q}\sim \rho_*$, where $\rho_*\equiv
\theta_*/\theta_{\rm E}$ is the angular size of the source $\theta_*$
in units of the angular Einstein ring radius $\theta_{\rm E}$.  The
Bayesian analysis presented in \S\ref{sec:micro} predicts a source
size in units of the Einstein ring radius of $\rho \simeq 10^{-2}$,
and therefore we expect perturbations from mass ratios of $\la
10^{-4}$ to be suppressed by finite source effects.  For the lowest
mass ratio we consider, $q=10^{-5}$, these suppressions are
significant and so we include finite source effects assuming $\rho =
10^{-2}$.

\begin{figure}[ht*]
\epsscale{1.0}
\plotone{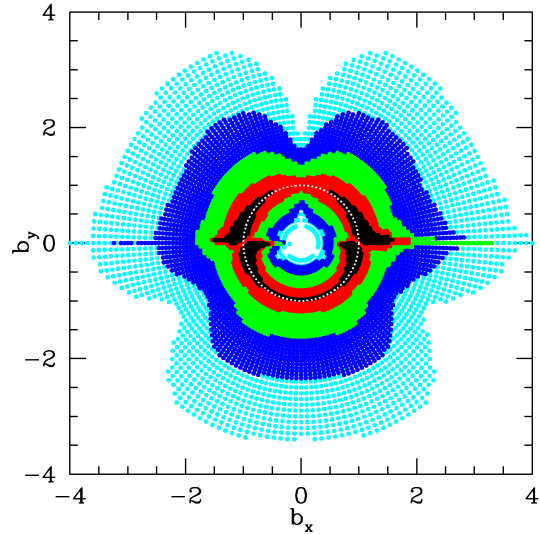}
\caption{
Planetary detection efficiency 
for simulated data of the GSC 3656-1328 microlensing event,
assuming that it was monitored densely over the peak
with a photometric precision of $\sim 0.4\%$.   
The points show the projected coordinates of the planet ($b_x,b_y$) relative
to the position of the primary star, at which the
the planet would be detected with $\Delta\chi^2\ge 160$.  Here $b_x$ is 
parallel to the direction of the relative source-lens proper motion.
The projected positions are in units of the Einstein ring radius of the primary;
the most likely value for the Einstein radius is $\sim 0.4~{\rm AU}$. 
The colors correspond to planet/star mass ratios of $10^{-3}$ (cyan), 
$10^{-3.5}$ (blue), $10^{-4}$ (green), $10^{-4.5}$ (red), 
and $10^{-5}$ (black). A mass ratio of
$\sim 10^{-4.5}$ corresponds to an Earth-mass companion
for the most likely primary mass of $\sim 0.1~M_\odot$.
\label{fig:chi2}
}
\end{figure}

Figure \ref{fig:chi2} shows the projected positions of the planet
relative to the position of the primary, where the planet would be
detected with $\Delta\chi^2\ge 160$. Here the two components of the
position of the planet ($b_x,b_y$) are in units of $R_{\rm E}$, and
$b_x$ is parallel to the direction of the relative source-lens proper
motion, such that the source moves from left to right relative to the
primary along a trajectory with $b_y=u_0=0.0237$.  The Bayesian
analysis predicts a primary mass of $M\sim 0.1~M_{\odot}$ and $R_{\rm
E} \sim 0.4~{\rm AU}$.  Therefore, a Neptune mass ($q\sim 10^{-3.3}$)
planet would be detectable over a range of projected separations of
$\sim 0.2-1~{\rm AU}$, and planets with masses as low as $\sim
0.3~M_{\oplus}$ ($q\sim 10^{-5}$) would be detectable for
some separations near the Einstein ring radius ($\sim 0.4~{\rm
AU}$).

Identifying all of the microlensing events within $4~{\rm kpc}$ would
require an all-sky (or at least all-Galactic-plane) survey that
reached a flux level roughly 60 times fainter than GSC 3656-1328,
or $V\sim 16$, 
to compensate partly for the greater distance and partly for the
increased interstellar extinction toward more distant targets.  It
would be straightforward to place 120 10 cm lenses, each backed by a
20 megapixel camera with 7'' pixels, on one single-axis mount,
and so continuously monitor to the required depth the $10,000\,\rm
deg^2$ that are within $60^\circ$ of the zenith at any given time.  To
cover the whole Galactic plane would require two such devices, one in
each hemisphere.  Such ``fly's eye'' telescopes would have heritage in
ongoing experiments such as ASAS \citep{pojmanski04} and SuperWASP
\citep{pollacco06} and could be considered the ``next generation''
successors to these experiments. In addition to detecting nearby
microlensing events, these telescopes would have a very large number
of other science applications.  Although the technical and survey requirements
for each application are fairly distinct, these telescopes could in principle be
used to detect thousands of planets as they transit their host star,
rapidly identify gamma-ray burst afterglows (in plenty of time to
alert gamma-ray satellites -- the reverse of the usual procedure), and
provide several days' advance warning for Tunguska-size meteors that
are expected to hit the Earth of order once per century with 10
megaton impacts \citep{pac06}.

\acknowledgments
We thank Weidong Li for assistance with the {\it Swift} UVOT data, Eric Ford
for help with MCMC, Kris Stanek for reminding us about \citet{pac95},
Martin Dominik, Pascal Fouqu{\'e}, and Stefan Dieters for enlightening
discussions, and Arto Oksanen, Becky Enoch, Dave Messier, Arne Henden,
and Tut Campbell for contributions of data.  We would like to thank
the referee, J.P. Beaulieu, for a helpful report.  The Peters
Automated Infrared Imaging Telescope (PAIRITEL) is operated by the
Smithsonian Astrophysical Observatory (SAO) and was made possible by a
grant from the Harvard University Milton Fund, the camera loan from
the University of Virginia, and the continued support of the SAO and
UC Berkeley.  The PAIRITEL project is further supported by NASA/{\it Swift}
Guest Investigator Grant No. NNG06GH50G. We thank M. Skrutskie for his
continued support of the PAIRITEL project.  We acknowledge the use of
public data from the {\it Swift} data archive.  A.~G.\ and S.~D.\ were
supported by NSF grant AST 042758.  C.~B.\ would like to acknowledge
support from the Harvard Origins of Life Initiative.

\appendix

\section{Properties of the Source Star}\label{app:source}

GSC 3656-1328 (TYC 3656-1328-1) has a spectral type of A0V-A1V, and is located
in the constellation Cassiopeia ($\alpha=00^{\rm h}09^{\rm m}21.\hskip-2pt^{\rm s}99$, $\delta=+54^\circ39' 43.\hskip-2pt''8{\rm ~[J2000.0]}$; $l=116.\hskip-2pt^\circ8158, 
b=-7.\hskip-2pt^\circ7092$).  The Tycho catalog \citep{hog00} gives a proper motion
of $\mu_\alpha=-2.0 \pm 2.8~{\rm mas~yr^{-1}},\mu_\delta=-6.1 \pm 2.7~{\rm mas~yr^{-1}}$,
and $V_T=11.387\pm 0.076$ and $B_T=11.507 \pm 0.054$.  We convert the latter
to $V=11.376 \pm 0.076$ and $B=11.478 \pm 0.054$ using the standard 
{\it Hipparcos} \citep{esa97} transformation.  This $V$ magnitude
is consistent with our more precise inferred value of $V=11.39 \pm 0.01$, which we adopt here. 
The color of $B-V=0.102 \pm 0.093$ is consistent with, but less accurate than, 
the value of $B-V=0.19 \pm 0.01$ measured by \citet{mikolajewski06b}
when GSC 3656-1328 was $\sim 1$ mag above baseline.  Because
we see no evidence for chromaticity, we adopt the more precise
determination of \citet{mikolajewski06b} as the baseline color of the source.

We estimate the distance to the source using its color
and magnitude, assuming a dereddenned color and magnitude 
appropriate for its A0V-A1V spectral type inferred by \citet{mikolajewski06a}.  
Adopting the spectral type/color
calibration of \citet{kenyon95} we estimate $E(B-V)=0.16-0.19$,
and assuming $R_V=3.1$, we estimate a $V$-band
extinction of $A_V = 0.5-0.6$.  The apparent $V$ magnitude 
then implies a distance of $D = 960-1070~{\rm pc}$.  For definiteness,
we adopt $A_V=0.6$ and $D=1~{\rm kpc}$.  Using the observed $V-K_s$ of the source
at baseline gives reasonably consistent results; however uncertainties about the 
possible presence of systematics in the PAIRITEL data, as well as potential contamination
from light due to the lens itself, make these results less secure despite the longer
wavelength baseline.

We estimate the angular size of the source to be $\theta_*\simeq 10~\mu{\rm as}$,
based on the dereddened $(V-K)_0$ color and $V_0$ magnitude of the source and the 
color-surface brightness relations of \citet{kervella04}.  The uncertainty
in this estimate is not important for our purposes, but it is roughly a few percent
due to the uncertainty in $(V-K)_0$ and $V_0$.

\section{Bayesian Constraints on the Lens Star Properties}\label{app:lens}

In order to provide constraints on the properties of the lens giving rise
to the microlensing event in GSC 3656-1328, we perform a Bayesian analysis,
which naturally accounts for priors on the expected populations of microlensing events
toward the line of sight of  GSC 3656-1328, as well as the observed
constraints from the light curve and additional (external) information.
Our analysis is similar to that done for other microlensing events 
(see, e.g., \citealt{yoo04,dong06,dominik06,bennett06}); however, there are some 
particular nuances in the analysis of this particular event that motivate 
an in-depth discussion.

We construct prior distributions of microlensing event parameters
using simple models of the mass, density, and velocity distributions
of the lens stars.  We adopt double exponential models for the thin
and thick disks.  Our thin disk model is the same as that used by
\cite{han95,han03} with a scale length of $3.5~{\rm kpc}$ and scale
height of $325~{\rm pc}$, and our thick disk model has the same scale
length but a scale height of $1~{\rm kpc}$.  We adopt a
\citet{chabrier03} lognormal mass function with a peak at
$M=0.079~M_\odot$ and a width of $0.69$ dex for stars with $M\le
M_\odot$ and the present day mass function as measured by \citet{reid02} with a logarithmic slope
of $\alpha=-4.2$ for stars with $M\ge M_\odot$.  We add remnants following \citet{gould00}.  We
adopt Gaussian distributions for the lens velocities that are
independent of position along the line of sight.  These have
means in the $U,V,W$ directions of $({\bar v}_U,{\bar v}_V,{\bar v}_W)=(0,214,0)~{\rm km~s^{-1}}$,  and 
one-dimensional velocity dispersions of
$(\sigma_U,\sigma_V,\sigma_W)=(35,28,35)~{\rm km~s^{-1}}$ for the thin
disk and twice this for the thick disk.

In the absence of external constraints, the likelihood of
a given lens mass, distance, and velocity, is just 
the contribution to the microlensing event rate, which for fixed
source proper motion and distance is given by,
\begin{equation}
{\cal L}_\Gamma \propto \frac{R_{\rm E} v_\perp}{M} \left(\rho_1 G_{U,1}G_{V,1}G_{W,1}+
\rho_2 G_{U,2}G_{V,2}G_{W,2}\right) f_X \left(\frac{dn}{dM}\right)_{X}
\end{equation}
Here subscripts ``1'' and  ``2'' are quantities for the thin and
thick disks, respectively, $\rho$ is the mass density at the position
of the lens,  and $G$ is the exponential velocity distribution.  For example,
\begin{equation}
G_{U,1}=\frac{1}{\sqrt{2\pi}\sigma_{U,1}}\exp\left(\frac{-[v_U-{\bar v}_U]^2}{2\sigma_{U,1}^2}\right),
\label{eqn:gu1}
\end{equation}
and similarly for the other distributions. The (normalized) mass function of each population of 
lenses (stars, brown dwarfs, etc) is given by $(dn/dM)_X$, and 
$f_X$ is the relative contribution to the number density from 
each type of lens. These are $66.3\%$, $29.7\%$, $3.4\%$, and $0.5\%$ for main-sequence
stars+brown dwarfs, white dwarfs, neutron stars, and black holes, respectively.  The lens velocity
relative to the observer-source line of sight is given by
\begin{equation}
{\bf v}_\perp = {\bf v}_{L,\perp}- {\bf v}_{O,\perp} \left(1-\frac{D_L}{D_S}\right) - {\bf v}_{S,\perp} \frac{D_L}{D_S}.
\label{eqn:vperp}
\end{equation}
Here ${\bf v}_{L,\perp}$, ${\bf v}_{S,\perp}$ and ${\bf v}_{O,\perp}$ are the projected velocities of the lens
and the observer perpendicular to observer-source line of sight, where
${\bf v}_{O,\perp}$ accounts for both the velocity of the Sun and the velocity
of the Earth at the time of the event.

The projected velocity of the 
source is just its  proper motion times its distance, both of which
are constrained (\S\ref{app:source}).  We account for these constraints
by including a term in the overall likelihood of the form
\begin{equation}
{\cal L}_{\mu} \propto \exp\left\{-\frac{1}{2}\left(\frac{\mu_\alpha-[-2.0~{\rm mas~yr^{-1}}]}{2.8~{\rm mas~yr^{-1}}}\right)^2\right\}
\exp\left\{-\frac{1}{2}\left(\frac{\mu_\delta-[-6.1~{\rm mas~yr^{-1}}]}{2.7~{\rm mas~yr^{-1}}}\right)^2\right\}.
\label{eqn:lmu}
\end{equation}
We fix the lens
distance at $D_S=1~{\rm kpc}$, but test the effects of changing this value.

We can account for the constraints from the light curve by including additional
likelihood terms.  We consider constraints from the fitted timescale of the event,
the blend flux in the $V$ band, the apparent lack of finite-source effects,
and, in some cases, the blend fluxes in the $JHK_s$ bands.  The constraint on the timescale
takes the form,
\begin{equation}
{\cal L}_{t_{\rm E}} \propto 
\exp\left[-\frac{1}{2}\left(\frac{t_{\rm E}-7.19~{\rm days}}{0.03~{\rm days}}\right)^2\right].
\label{eqn:lte}
\end{equation}
The light curve exhibits
marginal evidence of finite-source effects, with $\rho_*=0.032$ and an
upper limit of $\rho_*<0.044$ at the $3\sigma$ level.  
We note that from the observed color and flux of the source, $\theta_* = 10\,\mu$as (see \S\ref{app:source}), this limit on $\rho_*$ constrains the source-lens relative proper
motion to be $\mu = \theta_{\rm E}/t_{\rm E} >
10~{\rm mas~yr^{-1}}$.  Since the angular speed of the source is
$6.4\pm 3.9~{\rm mas~yr^{-1}}$ and that of the lens is expected to be similar,
this range is plausible. This also implies that the lack of finite source
effects does not provide a strong constraint on the properties of the lens,
other than ruling out lenses that are very close to the source.  Nevertheless,
we include this constraint 
on $\rho_*$ from the light curve by first determining $\Delta\chi^2_{fs}(\rho_*)$, the change in $\chi^2$ from the best-fit point-source
model as a function of $\rho_*$, when allowing all the other parameters to vary.  
The likelihood of a particular $\rho_*$ then takes the form,
\begin{equation}
{\cal L}_{\rho_*} \propto 
\exp\left[-\Delta\chi^2_{fs}(\rho_*)/2\right].
\label{eqn:lrhos}
\end{equation}

We also consider constraints on the flux of the lens arising from measurements
of (or limits on) the blend flux from the analysis of the light curve. We assume that brown dwarfs and
remnants are dark.  For main-sequence lenses we adopt mass-luminosity relations from the
solar-metallicity, 1 Gyr isochrone of \citet{siess00}, which in turn
adopt empirical transformations from effective temperature to 
standard filter luminosities from \citet{kenyon95}.  We apply a (small) correction
to convert the standard \citet{bessell88} $JHK$ magnitudes given in these isochrones to the 
2MASS system.\footnote{See http://www.astro.caltech.edu/$\sim$jmc/2mass/v3/transformations/.}
To estimate the extinction along the line of sight to the lens,
we assume a vertically exponential dust
disk with a scale height of $120~{\rm pc}$, normalized such that the total
extinction to the source is $A_V=0.6$ (see \S\ref{app:source}). 
We use a standard extinction law ($R_V=3.1$) to convert to other bandpasses.  
We then apply the constraints on the lens fluxes from the light curve,
assuming that the blend flux is entirely due to the lens star.  For the $V$-band
the strongest constraint on the blend flux is
$f_{V}=0.00058 \pm 0.00181$, where the units are such that $f_V=1$ corresponds to a $V=10$ star. 
If we are considering only the constraint from the $V$-band flux, then the likelihood is simply
\begin{equation}
{\cal L}_{lc} \propto \exp\left(-\Delta\chi^2_{lc}/2\right),
\label{eqn:lv}
\end{equation}
where $\Delta\chi^2=[(f_{V}-f_{V,{\rm model}})/\sigma_V]^2$, $\sigma_V=0.00181$,
and $f_{V,{\rm model}}$ is the $V$-band flux of the lens predicted by the model.
We can also include constraints from the blend fluxes in the infrared.  
The measured blend fluxes $f_J,f_H,f_{K_s}$ in the $JHK_s$ bands are given in \S\ref{sec:micro}.  Since
these fluxes are correlated with each other and with the $V$-band flux within the microlensing fit,
we must adopt a somewhat more sophisticated approach.  We first construct the vector ${\bf \Delta a}=(f_{V}-f_{V,{\rm model}},f_{J}-f_{J,{\rm model}},f_{H}-f_{H,{\rm model}},f_{K_s}-f_{K_s,{\rm model}})$,
where $f_{V,{\rm model}},f_{J,{\rm model}},...$ are the model fluxes.
From the microlensing fit to the light curve, we can also construct the covariance matrix ${\cal C}$ for
the four blend-flux parameters.   The 
difference in $\chi^2$ between the fluxes predicted by the model and the 
measured fluxes is
\begin{equation}
\Delta\chi^2_{lc} = \sum_{i=1}^{4}\sum_{j=1}^{4} \Delta a_i {\cal B}_{i,j} \Delta a_j,
\label{eqn:dchi2lc}
\end{equation}
where ${\cal B}\equiv {\cal C}^{-1}$.  
The likelihood is then determined using Equation (\ref{eqn:lv}), as before.

Finally, the likelihood of a particular parameter combination is given
by
\begin{equation}
{\cal L}_{tot}={\cal L}_\Gamma {\cal L}_{\mu} {\cal L}_{t_{\rm E}} {\cal L}_{\rho_*}
{\cal L}_{lc}.
\label{eqn:ltot}
\end{equation}

We construct {\it a posteriori} probability densities using the Markov chain Monte Carlo
method.  We first randomly choose values for the lens mass, distance, and
$(U,V,W)$ velocity components $v_U,v_V,v_W$, over a range of 
parameter space that is broad in comparison to
the posterior probability distributions. We also randomly choose whether 
the lens is a main-sequence star,
brown dwarf, white dwarf, neutron star, or black hole.  Finally,
we choose a random value for the two components of the proper motion
of the source.  We evaluate the relative likelihood of this parameter combination using
equation \ref{eqn:ltot}.  We then randomly move
to another point in the parameter space of lens mass, distance, velocity,
and source proper motion, and evaluate the likelihood of this
new parameter combination. We step in parameter space by adding to each
of the parameters a random value drawn from a
Gaussian distribution with zero mean, and dispersion chosen to efficiently sample
the likelihood surface.  Specifically, these dispersions are $0.043$ dex in $\log(M/M_\odot)$
for $M$,
$148~{\rm pc}$ for $D_L$, $\sigma_U,\sigma_V,\sigma_W$ for the lens
velocities, and $2.7~{\rm mas~yr^{-1}}$ for the two components of the source
proper motion.  If the ratio of the likelihood of the new parameter
combination to the old combination is greater than unity, we take the step.  Otherwise,
we draw a random number between 0 and 1. If this number is less than this
ratio of likelihoods, we take the step; otherwise it is rejected.  After discarding
the first $10^4$ steps, we record every $10^3$ steps until we form a chain of $10^5$
points.  We form 10 such chains, each starting from different initial conditions,
and calculate
the \citet{gelman92} $R$-statistic.  This is within $0.2\%$ of unity for
each parameter, indicating that the chains are well-mixed and converged.  We then
merge the chains and use the result for the final probability
distributions.  

We also test the effects of changing the source distance by $\pm
100~{\rm pc}$ and the $V$-band extinction to the source by $\pm
0.2~{\rm mag}$. We find that the medians of the probability
distributions change by $\la 5\%$ for all of the
parameters of interest.


\begin{thebibliography}{}

\bibitem[Afonso et al.(2006)]{afonso06} Afonso, C., et al.\ 
2006, \aap, 450, 233 

\bibitem[Alcock et al.(1993)]{alcock93} Alcock, C., et al.\ 
1993, \nat, 365, 621 

\bibitem[Alcock et al.(1996)]{alcock96} Alcock, C., et al.\ 
1996, \apj, 461, 84 

\bibitem[Alcock et al.(1997)]{alcock97} Alcock, C., et al.\ 
1997, \apjl, 491, L11 

\bibitem[Alcock et al.(2000)]{alcock00} Alcock, C., et al.\ 
2000, \apj, 542, 281 

\bibitem[Alcock et al.(2001)]{alcock01} Alcock, C., et al.\ 
2001, \nat, 414, 617 

\bibitem[Ansari et al.(1995)]{ansari95} Ansari, R., et al.\ 
1995, \aap, 299, L21 

\bibitem[Aubourg et al.(1993)]{aubourg93} Aubourg, E., et al.\ 
1993, \nat, 365, 623 

\bibitem[Beaulieu et al.(1995)]{beaulieu95} Beaulieu, J.~P., et 
al.\ 1995, \aap, 299, 168 

\bibitem[Beaulieu et al.(2006)]{beaulieu06} Beaulieu, J.-P., et 
al.\ 2006, \nat, 439, 437 

\bibitem[Bennett et al.(2006)]{bennett06} Bennett, D.~P., 
Anderson, J., \& Gaudi, B.~S.\ 2006, ApJ, submitted (astro-ph/0611448)

\bibitem[Bessell \& Brett(1988)]{bessell88} 
Bessell, M.~S., \&  Brett, J.~M.\ 1988, \pasp, 100, 1134 

\bibitem[Bloom et al.(2006)]{bloom06} Bloom, J.~S., Starr, 
D.~L., Blake, C.~H., Skrutskie, M.~F., \& Falco, E.~E.\ 2006, ASP 
Conf.~Ser.~351: Astronomical Data Analysis Software and Systems XV, 351, 
751 

\bibitem[Bond et al.(2004)]{bond04} Bond, I.~A., et al.\ 2004, 
\apjl, 606, L155 

\bibitem[Burrows et al.(2005)]{burrows05} Burrows, D.~N., et al.\ 
2005, Space Science Reviews, 120, 165 

\bibitem[Calchi Novati et al.(2005)]{calchi05} Calchi Novati, 
S., et al.\ 2005, \aap, 443, 911 

\bibitem[Chabrier(2003)]{chabrier03} Chabrier, G.\ 2003, \pasp, 
115, 763 

\bibitem[Chwolson(1924)]{chwolson24} Chwolson, O.\ 1924, 
Astronomische Nachrichten, 221, 329 

\bibitem[Colley \& Gott(1995)]{colley95} Colley, W.~N., \& Gott, 
J.~R.~I.\ 1995, \apj, 452, 82 

\bibitem[Cook et al.(1995)]{cook95} Cook, K.~H., et al.\ 1995, 
IAU Colloq.~155: Astrophysical Applications of Stellar Pulsation, 83, 221 

\bibitem[de Jong et al.(2004)]{dejong04} de Jong, J.~T.~A., et 
al.\ 2004, \aap, 417, 461 

\bibitem[della Valle(1994)]{dellavale94} della Valle, M.\ 1994, 
\aap, 287, L31 

\bibitem[della Valle \& Livio(1996)]{dellavale96} della Valle, M., 
\& Livio, M.\ 1996, \apjl, 457, L77 

\bibitem[Derue et al.(2001)]{derue01} Derue, F., et al.\ 2001, 
\aap, 373, 126 

\bibitem[Di Stefano(2005)]{distefano05} Di Stefano, R.\ 2005, preprint (astro-ph/0511633)

\bibitem[Di Stefano(2007)]{distefano07} Di Stefano, R.\ 2007, \apj, in press (arXiv:0712.3558)

\bibitem[Di Stefano(2008)]{distefano08} Di Stefano, R.\ 2008, \apjl, submitted (arXiv:0801.1511)

\bibitem[Di Stefano \& Night(2008)]{dinight08} Di Stefano, R.\ 2008, \apjl, in press (arXiv:0801.1510)


\bibitem[Dominik(2006)]{dominik06} Dominik, M.\ 2006, \mnras, 
367, 669 

\bibitem[Dong et al.(2006)]{dong06} Dong, S., et al.\ 2006, 
\apj, 642, 842 

\bibitem[Eddington(1920)]{eddington20} 
Eddington, A.S.\ 1920, Space, Time, and Gravitation (Cambridge: Cambridge
University Press), pp. 134 and 308

\bibitem[Einstein(1936)]{einstein36} Einstein, A.\ 1936, Science, 
84, 506 

\bibitem[ESA(1997)]{esa97} 
European Space Agency 1997, The {\it Hipparcos} and Tycho
Catalogues (SP-1200; Noordwijk: ESA)

\bibitem[Fukui et al.(2007)]{fukui07}
Fukui, A., et al.\ 2007, ApJ, 670, 423

\bibitem[Gehrels et al.(2004)]{gehrels04} Gehrels, N., et al.\ 
2004, \apj, 611, 1005 

\bibitem[Gelman \& Rubin(1992)]{gelman92}
Gelman, A., \& Rubin, D.B.\ 1992, Stat.\ Sci., 7, 457

\bibitem[Gould(2000)]{gould00} Gould, A.\ 2000, \apj, 535, 928 

\bibitem[Gould et al.(2004)]{gould04} Gould, A., Bennett, 
D.~P., \& Alves, D.~R.\ 2004, \apj, 614, 404 

\bibitem[Gould et al.(1994)]{gould94} Gould, A., 
Miralda-Escude, J., \& Bahcall, J.~N.\ 1994, \apjl, 423, L105 

\bibitem[Gould et al.(2006)]{gould06} Gould, A., et al.\ 2006, 
\apjl, 644, L37 

\bibitem[Hamadache et al.(2006)]{hamadache06} Hamadache, C., et 
al.\ 2006, \aap, 454, 185 

\bibitem[Han \& Gould(1995)]{han95} Han, C., \& Gould, A.\ 
1995, \apj, 447, 53 

\bibitem[Han \& Gould(2003)]{han03} Han, C., \& Gould, A.\ 
2003, \apj, 592, 172 

\bibitem[Han(2007)]{han07} 
Han, C.\ 2007, ApJ, submitted (arXiv:0708.1215)

\bibitem[H{\o}g et al.(2000)]{hog00} H{\o}g, E., et al.\ 
2000, \aap, 355, L27 

\bibitem[Kenyon \& Hartmann(1995)]{kenyon95} Kenyon, S.~J., \& 
Hartmann, L.\ 1995, \apjs, 101, 117 

\bibitem[Kervella et al.(2004)]{kervella04} Kervella, P., 
Th{\'e}venin, F., Di Folco, E., \& S{\'e}gransan, D.\ 2004, \aap, 426, 297 

\bibitem[Lasserre et al.(2000)]{lasserre00} Lasserre, T., et al.\ 
2000, \aap, 355, L39 

\bibitem[Liebes(1964)]{liebes64} Liebes, S.\ 1964, Physical 
Review , 133, 835 

\bibitem[Mikolajewski et al.(2006a)]{mikolajewski06a} Mikolajewski, M., 
Tomov, T., Niedzielski, A., \& Czart, K.\ 2006a, The Astronomer's Telegram, 
931, 1 

\bibitem[Mikolajewski et al.(2006b)]{mikolajewski06b} Mikolajewski, M., 
et al.\ 2006b, The Astronomer's Telegram, 943, 1 

\bibitem[Munari et al.(2006)]{munari06} Munari, U., Siviero, A., 
Tomasella, L., \& Valentini, M.\ 2006, Central Bureau Electronic Telegrams, 
718, 1

\bibitem[Nakano \& Tago(2006)]{nakano06} Nakano, S., \& Tago, 
A.\ 2006, Central Bureau Electronic Telegrams, 711, 1 

\bibitem[Nakano et al.(2006)]{nakano06b} Nakano, S., Kadota, K., 
Sakurai, Y., \& Waagen, E.\ 2006, Central Bureau Electronic Telegrams, 712, 
1

\bibitem[Nemiroff(1998)]{nemiroff98} Nemiroff, R.~J.\ 1998, \apj, 
509, 39 

\bibitem[Nemiroff \& Rafert(1999)]{nemiroff99} Nemiroff, R.~J., \& 
Rafert, J.~B.\ 1999, \pasp, 111, 886 

\bibitem[Paczy{\' n}ski(1986)]{pac86} Paczy{\' n}ski, B.\ 1986, \apj, 
304, 1 

\bibitem[Paczynski(1995)]{pac95} Paczynski, B.\ 1995, Acta 
Astronomica, 45, 345 

\bibitem[Paczynski(2006)]{pac06} 
Paczynski, B.\ 2006, \pasp, 118, 1621 

\bibitem[Palanque-Delabrouille et al.(1998)]{palanque98} 
Palanque-Delabrouille, N., et al.\ 1998, \aap, 332, 1 

\bibitem[Paulin-Henriksson et al.(2002)]{ph02} 
Paulin-Henriksson, S., et al.\ 2002, \apjl, 576, L121 

\bibitem[Pollacco et al.(2006)]{pollacco06} Pollacco, D.~L., et 
al.\ 2006, \pasp, 118, 1407 

\bibitem[Pojma{\'n}ski(2004)]{pojmanski04} Pojma{\'n}ski, G.\ 2004, 
Astronomische Nachrichten, 325, 553 

\bibitem[Reid et al.(2002)]{reid02} Reid, I.~N., Gizis, J.~E., 
\& Hawley, S.~L.\ 2002, \aj, 124, 2721 

\bibitem[Refsdal(1964)]{refsdal64} Refsdal, S.\ 1964, \mnras, 
128, 295 

\bibitem[Renn et al.(1997)]{renn97}
Renn, J., Sauer, T., \& Stachel, J. 1997, Science, 275, 184

\bibitem[Roming et al.(2005)]{roming05} Roming, P.~W.~A., et 
al.\ 2005, Space Science Reviews, 120, 95 

\bibitem[Samus \& Antipin(2006)]{samus06} Samus, N.~N., \& 
Antipin, S.~V.\ 2006, Central Bureau Electronic Telegrams, 718, 5 

\bibitem[Schneider et al.(1992)]{lenses} Schneider, P., 
Ehlers, J., \& Falco, E.~E.\ 1992, Gravitational Lenses (Berlin: Springer)

\bibitem[Siess et al.(2000)]{siess00} Siess, L., Dufour, E., \& 
Forestini, M.\ 2000, \aap, 358, 593 

\bibitem[Skillman \& Patterson(1993)]{skillman93} Skillman, D.~R., 
\& Patterson, J.\ 1993, \apj, 417, 298 

\bibitem[Skrutskie et al.(2006)]{skrutskie06} Skrutskie, M.~F., et 
al.\ 2006, \aj, 131, 1163 

\bibitem[Smak(1984)]{smak84} Smak, J.\ 1984, \pasp, 96, 5 

\bibitem[Smith(2003)]{smith03} Smith, M.~C.\ 2003, \mnras, 343, 
1172 

\bibitem[Soldner(1804)]{soldner04}
Soldner, J., 1804, {\it Berliner Astronomisches Jahrbuch f{\"{u}}r das Jahr 1804}, 161 

\bibitem[Spiegel et al.(2006)]{spiegel06} Spiegel, D., et al.\ 2006, The 
Astronomer's Telegram, 942, 1 

\bibitem[Thomas et al.(2005)]{thomas05} Thomas, C.~L., et al.\ 
2005, \apj, 631, 906 

\bibitem[Tisserand \& Milsztajn(2004)]{tisserand04}
Tisserand, P., \& Milsztajn, A.\ 2004, preprint (astro-ph/0501584)

\bibitem[Tisserand et al.(2007)]{tisserand07} Tisserand, P., et 
al.\ 2007, \aap, 469, 387 

\bibitem[Udalski et al.(1993)]{udalski93} Udalski, A., Szymanski, 
M., Kaluzny, J., Kubiak, M., Krzeminski, W., Mateo, M., Preston, G.~W., \& 
Paczynski, B.\ 1993, Acta Astronomica, 43, 289 

\bibitem[Udalski et al.(2000)]{udalski00} Udalski, A., Zebrun, 
K., Szymanski, M., Kubiak, M., Pietrzynski, G., Soszynski, I., \& Wozniak, 
P.\ 2000, Acta Astronomica, 50, 1 

\bibitem[Udalski et al.(2005)]{udalski05} Udalski, A., et al.\ 
2005, \apjl, 628, L109 

\bibitem[Uglesich et al.(2004)]{uglesich04} Uglesich, R.~R., 
Crotts, A.~P.~S., Baltz, E.~A., de Jong, J., Boyle, R.~P., \& Corbally, 
C.~J.\ 2004, \apj, 612, 877 

\bibitem[van Paradijs(1985)]{vp85} van Paradijs, J.\ 1985, 
\aap, 144, 199 

\bibitem[Yoo et al.(2004)]{yoo04} Yoo, J., et al.\ 2004, 
\apj, 616, 1204 

\end{thebibliography}
\end{document}